\documentclass[smallextended]{svjour3}%
\pdfoutput=1
\usepackage{fullpage}
\usepackage{amsmath}
\usepackage{amsfonts}
\usepackage{amssymb}
\usepackage{graphicx}%
\newcommand\blfootnote[1]{%
	\begingroup
	\renewcommand\thefootnote{}\footnote{#1}%
	\addtocounter{footnote}{-1}%
	\endgroup
}

\providecommand{\U}[1]{\protect\rule{.1in}{.1in}}

\journalname{Quantum Information Processing}
\begin{document}
	
	\title{Generalized concurrence measure for faithful quantification of multiparticle pure state entanglement using Lagrange's identity and wedge product}
	
	\titlerunning{Faithful quantification of multiparticle pure state entanglement}
	
	\author{Vineeth S. Bhaskara $^1$        \and
		Prasanta K. Panigrahi$^2$ 
	}
	
	\institute{ \at
		\email{bhaskaravineeth@gmail.com}
		\at
		\email{pprasanta@iiserkol.ac.in}\\ 
		$^1$ Indian Institute of Technology Guwahati, Guwahati 781 039, Assam, India\\
		$^2$ Indian Institute of Science Education And Research Kolkata, Mohanpur - 741 246, West Bengal, India}
	
	%\author{Vineeth S. Bhaskara}
	
	%\email[]{bhaskaravineeth@gmail.com}
	
	%\affiliation{Indian Institute of Technology Guwahati, Guwahati 781 039, Assam, India}
	%\author{Prasanta K. Panigrahi}
	%\email[]{pprasanta@iiserkol.ac.in}
	%\affiliation{Indian Institute of Science Education and Research Kolkata, Mohanpur 741 246, West Bengal, India}
	\date{}
	\maketitle

	\begin{abstract}
		Concurrence, introduced by Hill and Wootters [Phys. Rev. Lett. \textbf{78}, 5022 (1997)], provides an important measure of entanglement for a general pair of qubits that is faithful: strictly positive for entangled states and vanishing for all separable states. Such a measure captures the entire content of entanglement, providing necessary and sufficient conditions for separability. We present an extension of concurrence to multiparticle pure states in arbitrary dimensions by a new framework using the Lagrange's identity and wedge product representation of separability conditions, which coincides with the ``I-concurrence" of Rungta \textit{et al.} [Phys. Rev. A \textbf{64}, 042315 (2001)] who proposed by extending Wootters's spin-flip operator to a so-called universal inverter superoperator. Our framework exposes an inherent geometry of entanglement, and may be useful for the further extensions to mixed and continuous variable states.
		\keywords{Quantum entanglement \and Separability \and Multiparticle pure states \and Lagrange's identity \and Wedge product}
		\PACS{PACS 03.65.Ud \and PACS 03.67.-a \and PACS 02.10.Ud}
	\end{abstract}
	
	%\pacs{03.65.Ud, 03.67.-a, 02.10.Ud}
	%\keywords{}
	
	\maketitle
	
	
	\blfootnote{The final publication is available at Springer via http://dx.doi.org/10.1007/s11128-017-1568-0}
	
	\section{Introduction}
	\label{intro}
	
	A deeper understanding of inseparability or entanglement is of fundamental importance for the understanding of intrinsic quantum correlations. It has far reaching applications in quantum computation and information theory \cite{neilsen}. Entanglement forms an elementary resource in quantum computation and various quantum communication protocols \cite{key1,key2}. Detecting and quantifying this resource is of great practical application.

	Quantifying entanglement faithfully in a multiparticle scenario is central to quantum information theory so that one can estimate how close quantum states are to classical ones, and characterize the efficiency of protocols deterministically, which use entanglement as a resource~\cite{vlatko,Zhang2016,Tan16}. Recent interest on the connections between quantum entanglement and the emergence of space-time \cite{nature-qc-grav,raamsdonk} also calls for a systematic study of the geometry-entanglement relationship with the quantification of entanglement playing a subtler role in the context of quantum gravity.
	
	For the two-qubit case, an important measure of entanglement is the concurrence \cite{Wootters1}, which is strictly positive for entangled states, and vanishing for separable states. It provides the necessary and sufficient conditions of separability for a general pair of qubits. An extension of concurrence for multiparticle pure states is the ``I-concurrence" introduced by Rungta \textit{et al.} \cite{rungta}. They generalized the spin-flip superoperator to act on quantum systems of arbitrary dimensions, and introduced the corresponding generalized concurrence for joint pure states of bipartite quantum systems.
	
	In this paper, we present a similar generalization of concurrence to multiparticle pure states of arbitrary dimensions that is faithful by a new framework using the Lagrange's identity, and wedge product representation of the constraints amongst the complex amplitudes necessary for separability, leading to a measure of entanglement identical to the I-concurrence. This framework reveals an essential geometry of entanglement and may be useful for further extension of concurrence to other complex systems of interest.

	There have been works on a similar spirit, of which some include the study by Sawicki \textit{et al.} \cite{symplectic} on the symplectic geometry of entanglement, Nielsen \cite{major} on the connection between the algebra of majorization and entanglement transformations, Zhu \cite{Zhu16} on the structure of quantum correlations of many-body systems, Duan \textit{et al.} \cite{cirac} and Simon \cite{simonph} on the entanglement in continuous variable systems.

	%------------------------------------------
	
	\section{Separability for pure multiparticle states, and the central result}
	\label{sec:1}

	For future convenience, we define \textit{separability} for pure multiparticle states. Consider, a $n$-particle pure quantum system. Let $P|Q$ be a bipartition of this composite(whole) system $P\cup Q$, with respective Hilbert spaces $\mathcal{H}_P$ and $\mathcal{H}_Q$ for the states of the sub-systems $P$ and $Q$, then the state space of the composite system is given by the tensor product $\mathcal{H}=\mathcal{H}_P \otimes \mathcal{H}_Q$. If a pure state $|\psi\rangle \in \mathcal{H}$ of the composite system can be written in the form \[|\psi\rangle=|\phi\rangle_P \otimes |\chi\rangle_Q,\] where $|\phi\rangle_P \in \mathcal{H}_P$ and $|\chi\rangle_Q \in \mathcal{H}_Q$ are the pure states of the sub-systems $P$ and $Q$ respectively, then the system is said to be separable across the bipartition $P|Q$. Alternatively, the sub-system $P$ is separable from the composite system $P \cup Q$. Otherwise, the sub-systems $P$ and $Q$ are said to be entangled.

	To state the central result of the paper, consider a $n$-particle system with particles labelled by $(k)$, $k=1,2,...,n$. Suppose $|\psi\rangle$ is any pure state of the system and $\rho=|\psi\rangle\langle\psi|$ be its density matrix. Let $\mathcal{M}$ be the set of the \textit{particular} particles whose bipartite separability from the composite system is of interest with cardinality $m~(<n)$. Then the generalized concurrence, $E_{\mathcal{M}}$, for the bipartition $\mathcal{M}|~\overline{\mathcal{M}}$ is given, in equivalent forms, as ($\overline{\mathcal{M}}$ being the complementary set of $\mathcal{M}$): 
	
	\begin{align*}
	E_{\mathcal{M}}^2&=4~\sum_{i<j}\left(~\rho_{ii}^{\mathcal{M}}~\rho_{jj}^{\mathcal{M}}-\rho_{ij}^{\mathcal{M}}~\rho_{ji}^{\mathcal{M}}~\right)=4~\sum_{i<j} \lambda_i \lambda_j =2\bigg[1-  \operatorname{tr}\big[(\rho^{\mathcal{M}})^2\big]\bigg],  \end{align*}
	\\ where $\rho^{\mathcal{M}} \ \stackrel{\mathrm{def}}{=}\ \sum_j \langle j|_{\overline{\mathcal{M}}} \left( |\psi\rangle \langle\psi| \right) |j\rangle_{\overline{\mathcal{M}}} = \hbox{Tr}_{\overline{\mathcal{M}}} \; (\rho)$ is the reduced density matrix on the subsystem $\mathcal{M}$ obtained by tracing out the subsystem $\overline{\mathcal{M}}$,  and $\lambda_i$ are the eigenvalues of $\rho^{\mathcal{M}}$.
	
	%-----------here-------------
	
	$E_{\mathcal{M}}$ vanishes iff the system is separable across the bipartition $\mathcal{M}|~\overline{\mathcal{M}}$ and takes the maximum value iff $\rho^{\mathcal{M}}$ is maximally mixed. A measure of global entanglement would then be the sum of measures for distinct bipartitions of the system. 
	Evidently, a composite system is separable across all bipartitions if and only if every single-particle bipartition is separable. Therefore, the necessary and sufficient criterion for separability across all bipartitions is
	$ \sum_{k=1}^{n} E_{(k)}^2=0$,
	where $\rho^{(k)}$ is the single-particle reduced density matrix of $\rho$ on particle $(k)$.

	One can arrive at the result by considering the simple case of a two-qubit system, and subsequently generalizing the framework to multiparticle systems in arbitrary dimensions.
	
	\section{Two-qubit concurrence using Lagrange's identity and wedge product framework}
	
	Consider, a \emph{two-qubit system} with qubits $A$ and $B$. Let $|\psi\rangle$ be a normalized pure state of the system with \begin{equation*} |\psi\rangle = p|0_A0_B\rangle+q|0_A1_B\rangle+r|1_A0_B\rangle+s|1_A1_B\rangle \end{equation*} ($p,q,r,s \in \mathbb{C}$).
	Rewriting the state as: 
	\begin{align} |\psi\rangle&=|0_A\rangle~\big(p|0_B\rangle+q|1_B\rangle\big) + |1_A\rangle~\big(r|0_B\rangle+s|1_B\rangle\big)\\ &=|0_A\rangle~ \langle0_A|\psi\rangle+|1_A\rangle~\langle1_A|\psi\rangle ,  \nonumber\end{align}
	the bipartition $A|B$ is separable if and only if the vectors $\langle0_A|\psi\rangle=p|0_B\rangle+q|1_B\rangle$ and $\langle1_A|\psi\rangle=r|0_B\rangle+s|1_B\rangle$ (or, equivalently $\langle0_B|\psi\rangle$ and $\langle1_B|\psi\rangle$) are parallel, i.e., if and only if\begin{equation} \frac{p}{r} = \frac{q}{s} \Rightarrow ps-qr = 0.\end{equation} Then the modulus of $ps-qr$ is a faithful measure of entanglement for two qubits, which vanishes only for separable states. This condition may be elegantly written using the notation of a wedge product, which generalizes easily to multiparticle systems in arbitrary dimensions, as we show subsequently.
	
	In geometric algebra \cite{geoalgcam}, the \textit{wedge product} of two vectors is seen as a particular generalization of cross product to higher dimensions, and is defined as follows. Consider, any two vectors $\overrightarrow{a}$ and $\overrightarrow{b}$ in $\mathbb{C}^m$ written in the orthonormal basis $\{\hat{e}_i\}_{i=1}^{m}$. Their wedge product is a bivector in the $^mC_2$ dimensional exterior space with basis $\{\hat{e}_i\}_{i=1}^{m}\wedge\{\hat{e}_j\}_{j>i}^{m}$, defined by stipulating that $\hat{e}_{i}\wedge \hat{e}_{j}=-\hat{e}_j\wedge \hat{e}_i$ and $\hat{e}_{i}\wedge \hat{e}_{i}=0$, as:
	\begin{align}\overrightarrow{a} \wedge \overrightarrow{b}=\sum_{i=1}^{m-1} \sum_{j=i+1}^m \left(a_i b_j - a_j b_i\right)~ \hat{e}_i \wedge \hat{e}_j ,\end{align}
	with $\overrightarrow{a} \wedge \overrightarrow{a}=0$ and $\overrightarrow{a} \wedge \overrightarrow{b}=(-1)~\overrightarrow{b} \wedge \overrightarrow{a}$. In the coordinate notation $\overrightarrow{a} \wedge \overrightarrow{b}$ may be written as:	
	\begin{align*} (a_1b_2-a_2b_1,a_1b_3-a_3b_1,...,a_1b_m-a_mb_1,a_2b_3-a_3b_2,..., a_2b_m-a_mb_2,...,a_{m-1}b_m-a_mb_{m-1}) .\end{align*}

	This representation allows one to write the separability conditions in a compact and useful form.
	We note $||\langle0_A|\psi\rangle \wedge \langle1_A|\psi\rangle||=||\langle0_B|\psi\rangle \wedge \langle1_B|\psi\rangle||=|ps-qr|$, which is the measure of entanglement for the case of a two-qubit pure state. $||\langle0_A|\psi\rangle \wedge \langle1_A|\psi\rangle||$ geometrically represents the area of the complex parallelotope formed by the vectors $\langle0_A|\psi\rangle$ and $\langle1_A|\psi\rangle$ in the Hilbert space of qubit B. We write the two-qubit measure of entanglement as $E=2||\langle0_A|\psi\rangle\wedge\langle1_A|\psi\rangle||=2||\langle0_B|\psi\rangle\wedge\langle1_B|\psi\rangle||=2|ps-qr|,$ which is the concurrence~\cite{Wootters1} for two-qubit pure states defined by Hill and Wootters as
	$
	C(\psi)=|\langle\psi|\tilde{\psi}\rangle|=2|ps-qr|$, where $|\tilde{\psi}\rangle=\sigma_y |\psi^*\rangle$, $\sigma_y=
	\left( {\begin{array}{cc}
		0 & -i \\
		i & 0 \\
		\end{array} } \right)$ and $|\psi^*\rangle$ is the complex conjugate of $|\psi\rangle$.

	For maximal entanglement by this measure, the area of the parallelotope, $|ps-qr|$, must be maximum, which implies that the parallelotope must be a ``square" with its sides taking the maximum possible value. As the sum of the squares of the sides is constrained to be 1 (by normalization), i.e., $|\langle0_A|\psi\rangle|^2+|\langle1_A|\psi\rangle|^2=1$, the area is maximized when each of the side of the square equals $ \frac{1}{\sqrt{2}}$. Then, $E_{(max)}=1$, $0\leq E \leq1$. Therefore, for maximal entanglement, \begin{gather*}|\langle0_A|\psi\rangle|=|\langle1_A|\psi\rangle|=\frac{1}{\sqrt{2}}, |\langle0_B|\psi\rangle|=|\langle1_B|\psi\rangle|=\frac{1}{\sqrt{2}}, \\
	{(\langle0_A|\psi\rangle)}^\dagger\langle1_A|\psi\rangle=0, {(\langle0_B|\psi\rangle)}^\dagger\langle1_B|\psi\rangle=0 .\end{gather*}  These conditions for maximal entanglement are identical to the condition of the reduced density matrix being maximally mixed. 
	
	%--------------------------------

	Recall the generalized Lagrange's identity \cite{lars} for vectors in $\mathbb{C}^m$, which is a generalization of the Brahmagupta-Fibonacci identity \cite{brahma} and a special form of the Binet-Cauchy identity~\cite{linearalg,oliverpaper}. Consider, two vectors $\overrightarrow{a}, \overrightarrow{b} \in \mathbb{C}^m$. Then the Lagrange's identity takes the form: $ \|\overrightarrow{a}\|^2 \|\overrightarrow{b}\|^2- |\overrightarrow{a} \cdot \overrightarrow{b}|^2=\|\overrightarrow{a} \wedge \overrightarrow{b} \|^2 $ ($\| \cdot \|$ representing the norm of a vector, and  $|\cdot |$ the modulus of a scalar), i.e.,
	\begin{align}  \left( \sum_{k=1}^m |a_k|^2\right) \left(\sum_{k=1}^m |b_k|^2\right)-\left|\sum_{k=1}^m a_k \overline{b_k}\right|^2=\sum_{i=1}^{m-1} \sum_{j=i+1}^m |a_i b_j - a_j b_i|^2 \label{lagrange}
	\end{align} 
	where $\overline{b_k}$ represents the complex conjugate of $b_k$ (see Appendix A for proof). The norm of the wedge product $\overrightarrow{a} \wedge \overrightarrow{b}$ calculated by RHS of Eq. \eqref{lagrange} takes $\mathcal{O}(m^2)$ steps, while calculating the same using the LHS takes only $\mathcal{O}(m)$ steps. Therefore, this identity when applied to the wedge product representation of the separability conditions results in a computationally lesser intensive expression, asymptotically with increasing number of particles and dimensions, in terms of the traces of the squared reduced density matrices of the pure state. 
	
	By this identity, one may write $E_A^2=4||\langle0_A|\psi\rangle\wedge\langle1_A|\psi\rangle||^2=4||(p,q) \wedge (r,s)||^2$ as $\bigg[4(|p|^2+|q|^2)(|r|^2+|s|^2)-4|p\overline{r}+q\overline{s}|^2\bigg]$. By noting this to be the \textit{determinant} of the reduced density matrix on qubit A ($\rho^{A}$), by definition, as $\rho$ in this case is 
	\begin{align*} \rho = \left( {\begin{array}{cccc}
		|p|^2 & p\overline{q} & p\overline{r} & p\overline{s} \\
		q\overline{p} & |q|^2 & q\overline{r} & q\overline{s} \\
		r\overline{p} & r\overline{q} & |r|^2 & r\overline{s} \\
		s\overline{p} & s\overline{q} & s\overline{r} & |s|^2 \\
		\end{array} }\right), \end{align*}
	and therefore the reduced density matrix on A, $\rho^A$, takes the form:
	\begin{align*}
	\rho^A &= \langle 0_B|\rho|0_B\rangle + \langle 1_B|\rho|1_B\rangle \\
	&= \left( {\begin{array}{cc}
		|p|^2+|q|^2 & p \overline{r} + q \overline{s} \\
		r \overline{p} + s \overline{q} & |r|^2 + |s|^2 \\
		\end{array}} \right), \end{align*}
	one may, thus, rewrite the two-qubit measure of entanglement as $E=2\sqrt{det(\rho^{A})}=2\sqrt{det(\rho^{B})}$. This may further be written as \begin{align}E_A^2&=4~det(\rho^{A})=4\sum_{i<j} \left(~\rho^{A}_{ii}\rho^{A}_{jj}-\rho^{A}_{ij}\rho^{A}_{ji}~\right) \nonumber \\
	&=4\left[ \frac{1}{2}\sum_{i,j} \left(~\rho^{A}_{ii}\rho^{A}_{jj}-\rho^{A}_{ij}\rho^{A}_{ji}~\right)\right] \nonumber \\
	&=4\left[ \frac{1}{2}\bigg([\operatorname{tr}(\rho^{A})]^2 - \operatorname{tr}[(\rho^{A})^2]\bigg)\right] \label{frustrum}\\
	&=2~\bigg[1-  \operatorname{tr}[(\rho^{A})^2]\bigg],\label{frustrun}\end{align}
	since the trace of a valid density matrix is unity and for any square matrix $M$, $\sum_{i,j} M_{ij} M_{ji} = \operatorname{tr}(M^2)$, and $\sum_{i,j} M_{ii} M_{jj} = \sum_i M_{ii} \sum_j M_{jj} = \operatorname{tr}(M)^2$.
	
	The characteristic polynomial of a $m\times m$ matrix $M$ in $t$ is given by:
	\begin{align*}  t^m - (\operatorname{tr} M) t^{m-1} + \frac{1}{2}~\bigg(\operatorname{tr}(M)^2 - \operatorname{tr}(M^2)\bigg)~t^{m-2}   
	+ \cdots + (-1)^m ~(\det M).\end{align*}
	So Eq. \eqref{frustrum} is the $t^{m-2}$ coefficient (except for a constant) of the characteristic polynomial of the $m\times m$ reduced density matrix $\rho^A$. This can be thought of as the first step, interpolating between the trace of $\rho^A$ (which is the $t^{m-1}$ coefficient) and the determinant of $\rho^A$ (which is the constant coefficient). The roots of the characteristic polynomial are precisely the eigenvalues of $\rho^A$. If the eigenvalues of $\rho^A$ are $\lambda_1, \ldots, \lambda_m$ then ~\cite{vieta}
	\begin{align} E_A^2&=4\left[\frac{1}{2}~\left(1-\sum_i \lambda_i^2\right)\right]=4~\sum_{i<j} \lambda_i \lambda_j. \label{frustrup}
	\end{align}
	This mathematical setting extends in a straightforward way to more general cases in higher dimensions, and a global faithful measure of entanglement may be written down by summing over the contribution of each of the independent bipartitions of the general pure state as we show subsequently.
	
	% Note that $det(\rho)\geq0$, since a valid density matrix must necessarily be a positive semi-definite operator. 
	
	%Evidently, this measure is invariant under local unitary transformations, as it does not alter the determinant of the %reduced density matrix.
	
	\section{Extension to multiparticle states in arbitrary dimensions}

	Consider a $n$-particle pure state $|\psi\rangle$ in arbitrary dimensions with the particles labeled by \{$1,~2,~\ldots,~n$\} in an orthonormal basis as
	\begin{equation}
	|\psi\rangle = \sum_{j_1,~j_2,~...,~j_n=0}^{d_1 -1,~d_2 -1,~...,~d_n-1} a_{j_1j_2...j_n} ~~|j_1\rangle \otimes |j_2\rangle \otimes \cdots \otimes |j_n\rangle, \label{gen1}
	\end{equation}
	where particle $i$ has access to a $d_i$ dimensional Hilbert space, and $a_{j_1j_2...j_n}$ are the complex amplitudes. That is, particle $i$, when isolated, may be described by $d_i$ discrete orthonormal basis set \{$|0\rangle,|1\rangle,...,|d_i-1\rangle$\}. Therefore, $|\psi\rangle$ exists in a $D$-dimensional Hilbert space where $D=\prod_{i=1}^n d_i$. For convenience, one might omit the upper limits of the summation in Eq. \eqref{gen1} by noting that each summation index $j_i$  appropriately goes from $0$ to $d_i-1$. 
	
	Consider the bipartite separability of a particular set $\mathcal{M}$ of $m$-particles $(m<n)$ out of the $n$-particle system. Without any loss of generality, let the $m$-particles be labeled by \{$1,2,...,m$\}, so that the particles labeled by \{$m+1,m+2,...,n$\} represent the rest of $(n-m)$-particles belonging to the complement set $\overline{\mathcal{M}}$. One may rewrite the state $|\psi\rangle$ as
	\begin{align}
	|\psi\rangle &= \sum_{j_1,~j_2,~...,~j_n} a_{j_1j_2...j_n} ~~(|j_1\rangle \otimes |j_2\rangle \otimes \cdots \otimes |j_m\rangle)\otimes(|j_{m+1}\rangle \otimes |j_{m+2}\rangle \otimes \cdots \otimes |j_n\rangle) \nonumber \\
	&= \sum_{j_1,~j_2,~...,~j_m} ~~\sum_{j_{m+1},~...,~j_n} a_{j_1j_2...j_mj_{m+1}...j_n} ~~|j_1j_2...j_m\rangle\otimes |j_{m+1}...j_n\rangle \nonumber \\
	&=\sum_{j_1,~j_2,~...,~j_m}~\left[ |j_1j_2...j_m\rangle\otimes \left(\sum_{j_{m+1},~...,~j_n} ~ a_{j_1j_2...j_mj_{m+1}...j_n} |j_{m+1}...j_n\rangle \right) \right]. \label{gen2}
	\end{align}
	By noting that 
	\begin{equation} 
	\langle k_1k_2...k_m|\psi\rangle = \sum_{j_{m+1},...,j_n}~a_{k_1k_2...k_mj_{m+1}...j_n} ~ |j_{m+1}...j_n\rangle, \label{gen3}
	\end{equation}
	Eq. \eqref{gen2} may be expressed as
	\begin{equation}
	|\psi\rangle = \sum_{j_1,~...,~j_m}~ |j_1j_2...j_m\rangle~\otimes ~\Big[ \langle j_1j_2...j_m|\psi\rangle \Big].
	\end{equation}
	Therefore, for the separability of $|\psi\rangle$ across $\mathcal{M}|\overline{\mathcal{M}}$ bipartition, one needs the set of vectors $\{\langle j_1j_2...j_m|\psi\rangle\}_{j_1,...,j_m}$ in $\mathbb{C}^{D_{n-m}}$ (where $D_{n-m}=\prod_{i=m+1}^{n} ~d_i$) to be parallel for the $m$-particle state to factor out of $|\psi\rangle$. Therefore, the mutual wedge products among $\{\langle j_1j_2...j_m|\psi\rangle\}_{j_1,...,j_m}$ must vanish for the required bipartite separability. This is a necessary and sufficient condition of separability across $\mathcal{M}|\overline{\mathcal{M}}$ as noted before. Hence, one may construct a faithful measure of entanglement across the bipartition as
	\begin{equation}
	E_{\mathcal{M}}^2 = 4\sum_{i_1,...,i_m}~~\sum_{\substack{j_1\geq i_1,...,j_m\geq i_m \\|i_1-j_1|+...+|i_m-j_m| \neq 0}}~~ ||\langle i_1i_2...i_m|\psi\rangle \wedge \langle j_1j_2...j_m|\psi\rangle ||^2, \label{gen4}
	\end{equation}
	where the norm is computed in the orthogonal basis $\{|k_{m+1}...k_n\rangle \wedge |l_{m+1}...l_n\rangle\}_{l_{m+1}\geq k_{m+1},...,l_n\geq k_n}$, and $|i_1-j_1|+...+|i_m-j_m| \neq 0$ ensures that not all $i_p$ are equal to $j_p$ simultaneously $\forall~p$ in which case the wedge product trivially vanishes. There are $^{D_m} C_2$ terms in the above summation where $D_m = \prod_{i=1}^{m} d_i$. Noting that $|k_{m+1}...k_n\rangle \wedge |l_{m+1}...l_n\rangle = - |l_{m+1}...l_n\rangle \wedge |k_{m+1}...k_n\rangle$ by definition, consider
	\begin{align*}
	&\langle i_1i_2...i_m|\psi\rangle \wedge \langle j_1j_2...j_m|\psi\rangle \\
	&= \left(\sum_{k_{m+1},...,k_n} ~a_{i_1i_2...i_mk_{m+1}...k_n}~|k_{m+1}...k_n\rangle\right) \wedge \left(\sum_{l_{m+1},...,l_n} ~a_{j_1j_2...j_ml_{m+1}...l_n}~|l_{m+1}...l_n\rangle\right)  \\
	&=\sum_{k_{m+1},...,k_n}\sum_{l_{m+1},...,l_n} ~a_{i_1i_2...i_mk_{m+1}...k_n}a_{j_1j_2...j_ml_{m+1}...l_n}~|k_{m+1}...k_n\rangle \wedge |l_{m+1}...l_n\rangle  \\
	&= \sum_{k_{m+1},...,k_n}\sum_{\substack{l_{m+1}\geq k_{m+1},...,l_n\geq k_n \\ |i_{m+1}-k_{m+1}|+...+|l_n-k_n|\neq 0}} \\&\left( a_{i_1i_2...i_mk_{m+1}...k_n}a_{j_1j_2...j_ml_{m+1}...l_n} - a_{i_1i_2...i_ml_{m+1}...l_n}a_{j_1j_2...j_mk_{m+1}...k_n} \right) |k_{m+1}...k_n\rangle \wedge |l_{m+1}...l_n\rangle.
	\end{align*}
	Therefore,
	\begin{align*}
	& ||\langle i_1i_2...i_m|\psi\rangle \wedge \langle j_1j_2...j_m|\psi\rangle||^2 \\
	&= \sum_{k_{m+1},...,k_n}\sum_{\substack{l_{m+1}\geq k_{m+1},...,l_n\geq k_n \\ |i_{m+1}-k_{m+1}|+...+|l_n-k_n|\neq 0}} \left| a_{i_1i_2...i_mk_{m+1}...k_n}a_{j_1j_2...j_ml_{m+1}...l_n} - a_{i_1i_2...i_ml_{m+1}...l_n}a_{j_1j_2...j_mk_{m+1}...k_n} \right|^2.
	\end{align*}
	By the generalized Lagrange's identity Eq. \eqref{lagrange}, one may write the above expression equivalently as
	\begin{align*}
	=\left( \sum_{k_{m+1},...,k_n} \left| a_{i_1i_2...i_mk_{m+1}...k_n} \right|^2 \right)\left( \sum_{l_{m+1},...,l_n} \left| a_{j_1j_2...j_ml_{m+1}...l_n} \right|^2 \right) - \left|  \sum_{k_{m+1},...,k_n} \left( a_{i_1i_2...i_mk_{m+1}...k_n}\overline{a}_{j_1j_2...j_mk_{m+1}...k_n}  \right)  \right|^2.
	\end{align*}
	Hence, the entanglement measure may be explicitly written in terms of the amplitudes of the wavefunction in equivalent forms as \\ $E^2_{\mathcal{M}}$
	\begin{align*}
	=4\sum_{i_1,...,i_m}\sum_{\substack{j_1\geq i_1,...,j_m\geq i_m \\|i_1-j_1|+...+|i_m-j_m| \neq 0}} &\sum_{k_{m+1},...,k_n}\sum_{\substack{l_{m+1}\geq k_{m+1},...,l_n\geq k_n \\ |i_{m+1}-k_{m+1}|+...+|l_n-k_n|\neq 0}} \\ &\left| a_{i_1i_2...i_mk_{m+1}...k_n}a_{j_1j_2...j_ml_{m+1}...l_n} - a_{i_1i_2...i_ml_{m+1}...l_n}a_{j_1j_2...j_mk_{m+1}...k_n} \right|^2
	\end{align*}
	
	\begin{align}
	=4\sum_{i_1,...,i_m}\sum_{\substack{j_1\geq i_1,...,j_m\geq i_m \\|i_1-j_1|+...+|i_m-j_m| \neq 0}} \bigg[ \left( \sum_{k_{m+1},...,k_n} \left| a_{i_1i_2...i_mk_{m+1}...k_n} \right|^2 \right) &\left( \sum_{l_{m+1},...,l_n} \left| a_{j_1j_2...j_ml_{m+1}...l_n} \right|^2 \right)  \nonumber \\  &- \left|  \sum_{k_{m+1},...,k_n}  a_{i_1i_2...i_mk_{m+1}...k_n}\overline{a}_{j_1j_2...j_mk_{m+1}...k_n}  \right|^2 \bigg]. \label{gen5}
	\end{align} 
	The measure $E_{\mathcal{M}}$ is constructed (with appropriate constants) so that it coincides with Wootters's concurrence for the case of a two-qubit system. 
	
	Considering the pure state density matrix of the system as
	\begin{align}
	\rho &= |\psi\rangle\langle\psi| \nonumber\\
	&= \left(\sum_{j_1,j_2,...,j_n}  a_{j_1j_2...j_n}~|j_1j_2...j_n\rangle\right) \left(\sum_{i_1,i_2,...,i_n}  \overline{a}_{i_1i_2...i_n}~\langle i_1i_2...i_n|\right) \nonumber\\
	&= \sum_{j_1,j_2,...,j_n} \sum_{i_1,i_2,...,i_n} a_{j_1j_2...j_n}\overline{a}_{i_1i_2...i_n} ~|j_1j_2...j_n\rangle \langle i_1i_2...i_n|, \label{gen6}
	\end{align}
	one may define the reduced density matrix $\rho_{\mathcal{M}}$ of $\mathcal{M}$ by tracing out $\overline{\mathcal{M}}$ as
	\begin{align}
	\rho^{\mathcal{M}} \ & \stackrel{\mathrm{def}}{=}\ \hbox{Tr}_{\overline{\mathcal{M}}} \; (\rho) = \sum_{k_{m+1},...,k_n} \langle k_{m+1}...k_n|\rho |k_{m+1}...k_n\rangle \nonumber \\
	&= \sum_{k_{m+1},...,k_n} \sum_{j_{1},...,j_n} \sum_{i_{1},...,i_n} a_{j_1...j_mj_{m+1}...j_n} \overline{a}_{i_1...i_mi_{m+1}...i_n} \langle k_{m+1}...k_n |j_1...j_mj_{m+1}...j_n \rangle \langle i_1...i_mi_{m+1}...i_n |k_{m+1}...k_n \rangle \nonumber \\
	&=\sum_{j_{1},...,j_n} \sum_{i_{1},...,i_n} \underbrace{ \left[{\sum_{k_{m+1},...,k_n}  a_{j_1...j_mk_{m+1}...k_n} \overline{a}_{i_1...i_mk_{m+1}...k_n} } \right]}_{\text{matrix element of the reduced density matrix}}  |j_1...j_m\rangle \langle i_1...i_m| \label{gen7} \\
	&=\sum_{j_{1},...,j_n} \sum_{i_{1},...,i_n}   \rho^{\mathcal{M}}_{ji} |j_1...j_m\rangle \langle i_1...i_m|, \nonumber
	\end{align}
	where $j$ and $i$ are the indices of the reduced density matrix. The matrix element at the index $ji$ is given by $\rho^{\mathcal{M}}_{ji}=\langle j_1...j_m|\rho^{\mathcal{M}}|i_1...i_m\rangle $. Therefore, one arrives at the result considering 
	\begin{align}
	& 2\left[1 - \hbox{Tr} \left[ \left(\rho^{\mathcal{M}} \right)^2 \right] \right] \nonumber\\
	& = 4~\sum_{i<j}\left(~\rho_{ii}^{\mathcal{M}}~\rho_{jj}^{\mathcal{M}}-\rho_{ij}^{\mathcal{M}}~\rho_{ji}^{\mathcal{M}}~\right) ~~~ \text{[by Eq. \eqref{frustrun}]},\nonumber \\
	&= 4\sum_{i_1,...,i_m}\sum_{\substack{j_1\geq i_1,...,j_m\geq i_m \\|i_1-j_1|+...+|i_m-j_m| \neq 0}} \left(\langle i_1...i_m|\rho^{\mathcal{M}}|i_1...i_m\rangle\langle j_1...j_m|\rho^{\mathcal{M}}|j_1...j_m\rangle-\langle i_1...i_m|\rho^{\mathcal{M}}|j_1...j_m\rangle\langle j_1...j_m|\rho^{\mathcal{M}}|i_1...i_m\rangle\right) \nonumber \\
	& = E_{\mathcal{M}}^2~~~\text{[from Eq. \eqref{gen5} and Eq. \eqref{gen7}]}. \nonumber
	\end{align} 
	Since $1/D_m \leq \hbox{Tr} \left[ \left(\rho^{\mathcal{M}} \right)^2 \right] \leq 1$ (where the minimum is achieved when $\rho_{\mathcal{M}}$ is maximally mixed), therefore, $0 \leq E^2_{\mathcal{M}} \leq 2- 2/D_m $. Maximal entanglement across $\mathcal{M}|\overline{\mathcal{M}}$ is attained with $E_{\mathcal{M}}=\sqrt{2- 2/D_m}$ iff $\rho_{\mathcal{M}}$ is maximally mixed, by this measure.
	We analyze the above construction for the cases of a three-qubit, four-qubit, and two-qutrit system to assess the generalization.
	\paragraph{Three-qubit states:} 
	%------------------------
	Consider the {three-qubit} case. Let a normalized pure state of the three-qubit system be $|\psi\rangle$ with density matrix $\rho=|\psi\rangle\langle\psi|$ and with qubits labelled by $A$, $B$ and $C$. Let 
	\begin{align*}|\psi\rangle&=p|0_A0_B0_C\rangle+q|0_A0_B1_C\rangle+r|0_A1_B0_C\rangle +s|0_A1_B1_C\rangle+t|1_A0_B0_C\rangle+u|1_A0_B1_C\rangle+v|1_A1_B0_C\rangle+w|1_A1_B1_C\rangle \\
	&= |0_A\rangle~\big[ p|0_B0_C\rangle+q|0_B1_C\rangle+r|1_B0_C\rangle +s|1_B1_C\rangle \big] + |1_A\rangle~\big[t|0_B0_C\rangle+u|0_B1_C\rangle+v|1_B0_C\rangle+w|1_B1_C\rangle\big] \\
	&=|0_A\rangle~ \langle0_A|\psi\rangle+|1_A\rangle~\langle1_A|\psi\rangle
	\end{align*} 
	($p,q,r,s,t,u,v,w \in \mathbb{C}$). Similar to the two-qubit case, for separability of qubit A (i.e., the bipartition $A|BC$) here, the vectors $\langle0_A|\psi\rangle, \langle1_A|\psi\rangle$ must be parallel. This yields the condition for separability of qubit A to be: \begin{equation} \frac{p}{t}=\frac{q}{u}=\frac{r}{v}=\frac{s}{w}.\label{eight} \end{equation} These separability constraints amongst the complex amplitudes may be written in the wedge product representation as $\langle0_A|\psi\rangle\wedge\langle1_A|\psi\rangle=0$, which is equivalent to the relations in Eq. \eqref{eight} on cross-multiplying, since: \begin{align*}(p,q,r,s)\wedge(t,u,v,w)=(pu-qt,pv-rt,pw-st,qv-ru,qw-su,rw-sv),\end{align*} by the coordinate notation of the wedge product defined previously. Therefore, the bipartite separability $A|BC$ $\Leftrightarrow$ $\langle0_A|\psi\rangle\wedge\langle1_A|\psi\rangle=0$. Hence, its norm is a deterministic  measure of entanglement of qubit $A$ with system $BC$. By the Lagrange's identity, $\|\langle0_A|\psi\rangle\wedge\langle1_A|\psi\rangle\|^2$ turns out to be equal to the determinant of qubit A's reduced density matrix $\rho^{A}$ by definition, similar to the previous case. Therefore, one can write the global measure of entanglement for a three-qubit system, considering independent bipartitions, as: \begin{align*}E&=E_A+E_B+E_C\\&=2||\langle0_A|\psi\rangle\wedge\langle1_A|\psi\rangle||+2||\langle0_B|\psi\rangle\wedge\langle1_B|\psi\rangle|| +2||\langle0_C|\psi\rangle\wedge\langle1_C|\psi\rangle||\\
	&=2 \left[\sqrt{det(\rho^{A})}+\sqrt{det(\rho^{B})}+\sqrt{det(\rho^{C})}\right].\end{align*} This can be rewritten in terms of eigenvalues of the reduced density matrices by the derivation in Eq. \eqref{frustrup}.
	The maximum norm of each of the wedge products above is $=\frac{1}{2}$. Therefore, $0\leq E\leq 3$. The $|GHZ\rangle_{3}$ state is maximally entangled three-qubit state with $E=3$, by this measure, and for the $|W\rangle_{3}$ state, $E=2\sqrt{2}\simeq2.828$, which suggests that it is highly entangled but lesser than $|GHZ\rangle_{3}$ state, where:
	$ |GHZ\rangle_{3}=\frac{1}{\sqrt{2}} \left(|0_A0_B0_C\rangle+|1_A1_B1_C\rangle \right)$ and
	$|W\rangle_3=\frac{1}{\sqrt{3}} \left(|1_A0_B0_C\rangle+|0_A1_B0_C\rangle+|0_A0_B1_C\rangle \right) .$
	
	Analogously, for a $n$-qubit system with pure state $|\psi\rangle$ and density operator $\rho$, separability of qubit labelled by $   ``i" ~(\leq n)$ $\Leftrightarrow$ $\langle0_i|\psi\rangle\wedge\langle1_i|\psi\rangle=0.$
	By Lagrange's identity this simplifies to: $det(\rho^{i})=0.$ Therefore, a particular qubit is separable from a $n$-qubit system if and only if its corresponding single-qubit reduced density matrix is singular. For the separability of the system across every bipartition, each single-qubit reduced density matrix being singular is necessary and sufficient.

	\paragraph{Four-qubit states:}
	Consider, a {four-qubit} system with qubits labelled by $A$, $B$, $C$ and $D$. Let $|\psi\rangle$ be its pure state with density matrix $\rho=|\psi\rangle\langle\psi|$. $E_A=2\|\langle0_A|\psi\rangle \wedge \langle1_A|\psi\rangle\|=2\sqrt{det(\rho^{A})}$ determines the separability of qubit A or qubit system (BCD) from the composite system (ABCD), similar to the previous cases. Analogous to the previous construction, for the separability of qubits (AB) or (CD) from the system, the vectors $\langle0_A0_B|\psi\rangle, \langle0_A1_B|\psi\rangle, \langle1_A0_B|\psi\rangle, \langle1_A1_B|\psi\rangle$ in the Hilbert space $\mathcal{H_{CD}}$ of qubit system (CD) must be parallel. This can be seen by writing $|\psi\rangle$ as \big[$|0_A0_B\rangle~\langle0_A0_B|\psi\rangle + |0_A1_B\rangle~\langle0_A1_B|\psi\rangle + |1_A0_B\rangle~\langle1_A0_B|\psi\rangle + |1_A1_B\rangle~\langle1_A1_B|\psi\rangle$\big]. Therefore, a non-vanishing wedge product of one of the vectors with any other among $\langle0_A0_B|\psi\rangle, \langle0_A1_B|\psi\rangle, \langle1_A0_B|\psi\rangle$ and $\langle1_A1_B|\psi\rangle$, indicates entanglement of the sub-systems (AB) and (CD). Therefore, define $E_{AB}$ as:
	\begin{align} E_{AB}^2=4~\big[~||\langle0_A&0_B|\psi\rangle\wedge\langle0_A1_B|\psi\rangle||^2+||\langle0_A0_B|\psi\rangle\wedge\langle1_A0_B|\psi\rangle||^2+||\langle0_A0_B|\psi\rangle\wedge\langle1_A1_B|\psi\rangle||^2+\nonumber\\ ||&\langle0_A1_B|\psi\rangle\wedge\langle1_A0_B|\psi\rangle||^2+||\langle0_A1_B|\psi\rangle\wedge\langle1_A1_B|\psi\rangle||^2+||\langle1_A0_B|\psi\rangle\wedge\langle1_A1_B|\psi\rangle||^2~\big]. \label{abey}\end{align}  
	Therefore, separability of bipartition $AB|CD$ $\Leftrightarrow$ $E_{AB}=0$. Again by the Lagrange's identity, the expression Eq. \eqref{abey} for $E_{AB}^2$ simplifies to the similar form as:
	\begin{align} E_{AB}^2=4\sum_{i,j=1, i<j}^{2^2} \left(\rho_{ii}^{AB}\rho_{jj}^{AB}-\rho_{ij}^{AB}\rho_{ji}^{AB}\right) \nonumber
	=4~\sum_{i<j} \lambda_i \lambda_j=2~\bigg[1-  \operatorname{tr}[(\rho^{AB})^2]\bigg],\nonumber\end{align}
	where $\lambda_i$ are the eigenvalues of $\rho^{AB}$. Note that the term $\sum_{i,j=1, i<j}^{2^2} \left(\rho_{ii}^{AB}\rho_{jj}^{AB}-\rho_{ij}^{AB}\rho_{ji}^{AB}\right)$ above is not the determinant of $\rho^{AB}$. Therefore, note that the generalizing expression is in terms of the traces of the squared reduced density matrices but not in terms of their determinants for general cases. Similar expressions follow for $E_{AC}^2$ and $E_{AD}^2$.
	Considering independent bipartitions one can write the global measure of entanglement for the four-qubit system as: 
	\begin{align*} E&=E_A+E_B+E_C+E_D+E_{AB}+E_{AC}+E_{AD}. \end{align*}
	Evidently, $E$ takes the maximum value only when the reduced density matrices are maximally mixed. Therefore, $E_{(max)}=4+\frac{3\sqrt{6}}{2}\simeq7.674$ for maximal entanglement, by this measure. But this may not be attained for the case of a four-qubit system, as shown by Higuchi \textit{et al.}~\cite{higuchi}. Therefore, $0\leq E<4+\frac{3\sqrt{6}}{2}$.

	For $|GHZ\rangle_{4}=\frac{1}{\sqrt{2}} \left(|0_A0_B0_C0_D\rangle+|1_A1_B1_C1_D\rangle \right)$ state, $E=7$, and for the four-qubit Higuchi-Sudbery state found numerically by Higuchi~\textit{et al.}~\cite{higuchi}:
	\begin{align*} |HS\rangle = \frac{1}{\sqrt{6}} [ |0011\rangle + |1100\rangle + \omega (|1010\rangle + |0101\rangle )  + \omega^2 (|1001\rangle + |0110\rangle) ], \end{align*} 
	where $\omega = e^{2\pi i/3}$, $E=4+2\sqrt{3}\simeq7.464$, which is close to the unattainable bound of $\simeq7.674$, showing that it is more entangled than the $|GHZ\rangle_4$ state, by this measure.

	\paragraph{Two-qutrit states:}
	Consider, a two-qutrit system with levels $|0\rangle,|1\rangle,|2\rangle$ and qutrits labeled by $A$ and $B$. Let $|\psi\rangle$ be its pure state and $\rho$ its density matrix. Similar to the previous reasoning, for separability of qutrit A, the vectors $\langle0_A|\psi\rangle,\langle1_A|\psi\rangle, \langle2_A|\psi\rangle$ must be parallel. This is clear once the state is written as: $|\psi\rangle=|0_A\rangle~(\langle0_A|\psi\rangle)+|1_A\rangle~(\langle1_A|\psi\rangle)+|2_A\rangle~(\langle2_A|\psi\rangle)$. Therefore, define the measure of entanglement of qutrit A with qutrit B as:
	\begin{align*} E_A^2=4~\big[||\langle0_A|\psi\rangle\wedge \langle1_A|\psi\rangle||^2+||\langle0_A|\psi\rangle\wedge\langle2_A|\psi\rangle||^2 +||\langle1_A|\psi\rangle\wedge\langle2_A|\psi\rangle||^2\big].
	\end{align*} Applying the Lagrange's identity gives (where $\lambda_i$ are the eigenvalues of $\rho^{A}$): \begin{align*}E_A^2=4\sum_{i,j=1, i<j}^{3}\left( \rho^{A}_{ii}~\rho^{A}_{jj}-\rho^{A}_{ij}~\rho^{A}_{ji}\right )
	=4\sum_{i<j} \lambda_i \lambda_j=2~\bigg[1-  \operatorname{tr}[(\rho^{A})^2]\bigg].\end{align*}

	One thus arrives at the result for pure multiparticle states in arbitrary dimensions (which also includes systems  of mixed dimensions like qubit-qutrit, among others) by noting the generalizing structure from the various cases above. A global measure of entanglement for the multiparticle system may be constructed by summing over the measures for distinct bipartitions of the system.
	
	\section{Conclusion}
	We hope our work provides new insights into the deeply interesting phenomenon of entanglement, exposing its essential geometry and mathematical structure, and is of relevance to various related problems like separability of mixed states and continuous variable systems, classification of  entanglement transformations, and entanglement characterization. This framework gives a faithful, computable measure of entanglement for pure states, and may further be useful in generalizing concurrence for mixed and continuous variable states. The measure may also be used in numerical searches for highly entangled multiparticle states ~\cite{brown,borras,mexicopoland}, without missing any useful state, to improve existing and discover new quantum information processing protocols~\cite{sre,Gao16}.

	%\section{Acknowledgments}
	\begin{acknowledgements}
		Bhaskara is thankful to Oliver Knill for pointing out the Binet-Cauchy identity. This work was supported by the National Initiative on Undergraduate Science (NIUS) undertaken by the Homi Bhabha Centre for Science Education, Tata Institute of Fundamental Research (HBCSE--TIFR), Mumbai, India. The authors acknowledge Vijay Singh and Praveen Pathak for continuous encouragement.
	\end{acknowledgements}

	\appendix
	\section{Proof of Lagrange's identity}Consider,
	\begin{align*}& RHS=||\overrightarrow{a}\wedge \overrightarrow{b}||^2\\&=\sum_{i=1}^{m-1} \sum_{j=i+1}^m |a_i b_j - a_j b_i|^2  \\
	&=\frac{1}{2} \sum_{i=1}^{m} \sum_{j=1}^{m} ~ |a_i b_j - a_j b_i|^2 \\
	&=\frac{1}{2} \sum_{i=1}^{m} \sum_{j=1}^{m} ~ (a_{i}b_{j}-a_{j}b_{i})(\overline{a}_{i}\overline{b}_{j}-\overline{a}_{j}\overline{b}_{i}) \\
	&=\frac{1}{2} \sum_{i=1}^{m} \sum_{j=1}^{m}~(|a_{i}|^2|b_{j}|^2-2Re(a_ib_j \overline{a}_{j} \overline{b}_{i})+|a_{j}|^2|b_{i}|^2) \\
	&=\left(\sum_{i=1}^m |a_i|^2\right)\left(\sum_{j=1}^m |b_j|^2\right)-Re\sum_{i=1}^m \sum_{j=1}^m (a_ib_j\overline{a}_{j} \overline{b}_{i}) \\ 
	&=\left(\sum_{i=1}^m |a_i|^2\right)\left(\sum_{j=1}^m |b_j|^2\right)- \left|\sum_{i=1}^m a_i \overline{b}_{i}\right|^2 \\
	&=\|\overrightarrow{a}\|^2 \|\overrightarrow{b}\|^2- |\overrightarrow{a} \cdot \overrightarrow{b}|^2=LHS.
	\end{align*} 
	Hence the identity.
	%\bibliography{references}

\end{document}